\documentclass[trackchanges,twocolumn]{aastex63}

\newcommand{\msolph}{\,$h^{-1}$M$_{\odot}$}
\newcommand{\Msun}{~h^{-1} \rm M_{\odot}}
\newcommand{\mpcph}{\,$h^{-1}$Mpc}

\shorttitle{Cosmic Voids as Massive Cluster Counterparts}
\shortauthors{Shim et al.}

\begin{document}

\title{Identification of Cosmic Voids as Massive Cluster Counterparts}



\author[0000-0001-7352-6175]{Junsup Shim}
\affiliation{School of Physics, Korea Institute for Advanced Study, 85 Hoegiro, Dongdaemun-gu, Seoul 02455, Korea}

\author{Changbom Park}
\affiliation{School of Physics, Korea Institute for Advanced Study, 85 Hoegiro, Dongdaemun-gu, Seoul 02455, Korea}

\author[0000-0002-4391-2275]{Juhan Kim}
\affiliation{Center for Advanced Computation, Korea Institute for Advanced Study, 85 Hoegiro, Dongdaemun-gu, Seoul 02455, Korea}

\author[0000-0003-3428-7612]{Ho Seong Hwang}
\affiliation{School of Physics, Korea Institute for Advanced Study, 85 Hoegiro, Dongdaemun-gu, Seoul 02455, Korea}
\affiliation{Korea Astronomy and Space Science Institute, 776 Daedeokdae-ro, Yuseong-gu, Daejeon 34055, Korea}

\begin{abstract}
We develop a method to identify cosmic voids from the matter density field by adopting a physically-motivated concept that voids are the counterpart of massive clusters. To prove the concept we use a pair of $\Lambda$CDM simulations, a reference and its initial density-inverted mirror simulation, and study the relation between the effective size of voids and the mass of corresponding clusters. Galaxy cluster-scale dark matter halos are identified in the Mirror simulation at $z=0$ by linking dark matter particles. The void corresponding to each cluster is defined in the Reference simulation as the region occupied by the member particles of the cluster. We study the voids corresponding to the halos more massive than $10^{13}\Msun$. We find a power-law scaling relation between the void size and the corresponding cluster mass. Voids with corresponding cluster mass above $10^{15}\Msun$ occupy $\sim1\%$ of the total simulated volume, whereas this fraction increases to $\sim54\%$ for voids with corresponding cluster mass above $10^{13}\Msun$. It is also found that the density profile of the identified voids follows a universal functional form. Based on these findings, we propose a method to identify cluster-counterpart voids directly from the matter density field without their mirror information by utilizing three parameters such as the smoothing scale, density threshold, and minimum core fraction. We recover voids corresponding to clusters more massive than $3\times10^{14}\Msun$ at 70--74 \% level of completeness and reliability. Our results suggest that we are able to identify voids in a way to associate them with clusters of a particular mass-scale.

\end{abstract}


\keywords{methods: data analysis -- methods: statistical -- large-scale structure of Universe}

\section{Introduction}\label{sec:Intro}
Cosmic voids are vast holes in the distribution of galaxies, which appear devoid of matter \citep{joeveer+78, kirshner+81, lapparent+86, geller&huchra89, shectman+96, hoyle&vogeley04, pan+12, krolewski+18}. Being underdense regions, voids experience gravity effectively pulling outward and are relatively more dominated by dark energy. Hence, void properties are intimately tied to the nature of gravity and dark energy. In addition, voids are less affected by non-linear processes \citep{kim&park98} and less contaminated by baryon physics compared to overdense regions. Thanks to these characteristics, voids have become a promising probe for testing cosmological models \citep[for a general overview on voids, see][]{weygaert&platen11}. For example, direct comparisons of void properties between observations and simulations have been conducted to test the $\Lambda$CDM cosmology \citep[e.g.][]{park+12,hwang+16}. In particular, the void statistics including abundance, shape, and density profile have been utilized to constrain dark energy \citep{lee&park09,lavaux&wandelt10,bos+12,pisani+15,verza+19}, the interaction between dark sectors \citep{li11,sutter+15}, gravity theories \citep{peebles01,nusser+05,li+12,cai+15}, and initial conditions \citep{blumenthal+92,weygaert&kampen93,colberg+05}.

Theoretically, a void is a well-defined structure in the spherical expansion model as the spherical underdense region with a high-density boundary surrounding it. The boundary of a void forms when the inner expanding mass catches up the outer mass at the moment of ``shell-crossing'' \citep{filmore&goldreich84, bertschinger85, blumenthal+92, suto+84,dubinski+93}. At the shell-crossing, the predicted spherically averaged density of a void is approximately $\delta_{\rm sc}\simeq-0.8$ \citep{blumenthal+92,dubinski+93,sheth&weygaert04}.

To define voids from simulation and observation, a variety of identification schemes have been proposed \citep[e.g.][]{plionis&basilakos02, hoyle&vogeley04, shandarin+04, colberg+05, brunino+07, hahn+07, platen&jones07, neyrinck08, lavaux&wandelt10, sousbie11, cautun+13, sutter+15}. However, it is uncertain whether voids identified using these algorithms represent theoretical voids predicted from the spherical expansion model. For example, \cite{jennings+13} showed that void abundance measured using $N-$body simulations significantly differs from that predicted in two-barrier excursion formalism (SvdW) \citep{sheth&weygaert04}. On the other hand, \cite{nadathur&hotchkiss15} and \cite{achitouv+15} found that voids identified with a watershed-based method exhibit a broad distribution of average densities, which is far from the shell-crossing density predicted in the spherical expansion model. In a similar context, it has been concluded that the shell-crossing density has to be modified so that the SvdW model matches the measured void abundances \citep{furlanetto&piran06,chan+14,sutter+14b}.

To make matters worse, dissimilar voids are identified when different void-finders are applied to a given density field \citep[for the comparison of identified voids, see][]{colberg+08}. This incompatibility may originate from the different assumptions on the shape, density, and dynamics of voids. The discrepancy between theoretical and identified voids, which also depends on the void finders, raises a fundamental question whether there is a physically motivated definition of voids that can be consistently identified in data, for example, from density fields. To address this issue, we take a new approach. We adopt the concept that a cosmic void is the counterpart of a massive cluster which is theoretically well defined and easily identified from simulations and observations.

According to the standard scenario of structure formation, clusters are expected to form at high peaks of the initial density field whereas voids are expected to form at low density troughs. Therefore, if the initial density fluctuations are inverted in sign, a cluster formation site will become a void formation site. Adopting the concept, at the Aspen cosmic void workshop held in 2006, one of the authors (CBP) proposed to use ``the mirror universe simulations'' to find the cluster -  void connection. He used a pair of simulations (the SCDM models) whose initial density fields had been swapped and presented the results that the voids corresponding to clusters more massive than $10^{14}\Msun$ occupied $9\%$ of mass, filled $27\%$ of the volume at $z=0$, and the underdensity of the void edge was more or less the same, $-0.6$. Recently, \citet{pontzen+16} independently revisited this void-halo duality by analyzing the relation between halos and ``anti''-halos in paired simulations with inverted initial conditions. (None of the authors of this paper was present at the Aspen Workshop.) They calculated the mean density within the Lagrangian volume occupied by each anti-halo and concluded that anti-structures of halos resemble voids. It was also shown that antihalo regions are well described by Zel'dovich approximation \citep{Stopyra+21}.

We adopt the concept and study the physical properties of cluster-counterpart voids, and develop a method to identify such voids from a density field. This paper is organized as follows. We describe our simulations and explain how we define voids as cluster counterparts in Section~\ref{sec:simulation}. In Section~\ref{sec:result}, the radial density profiles and size distribution of the voids are presented. We describe a method to identify the voids from a density field and summarize our results in Section~\ref{sec:discussion} and Section~\ref{sec:conclusion}, respectively.

\section{Methods}\label{sec:simulation}
\subsection{Mirror simulation}
We use a pair of the Multiverse simulations \citep{park+19,hong+20,tonegawa+20} to study the relation between voids and galaxy cluster-scale dark matter halos. The Multiverse simulations are a set of $N$-body simulations performed with the GOTPM \citep{dubinski+04} code varying the matter density parameter $\Omega_m$ and the equation of state of dark energy $w$. Each simulation has $N_p=2048^3$ particles in a cubic box of a side length $L_{\rm box}=1024~h^{-1}{{\rm Mpc}}$ which dictates the particle mass $M_p \simeq 9.02\times 10^9 \Msun$.

\begin{figure}
	\includegraphics[width=\columnwidth]{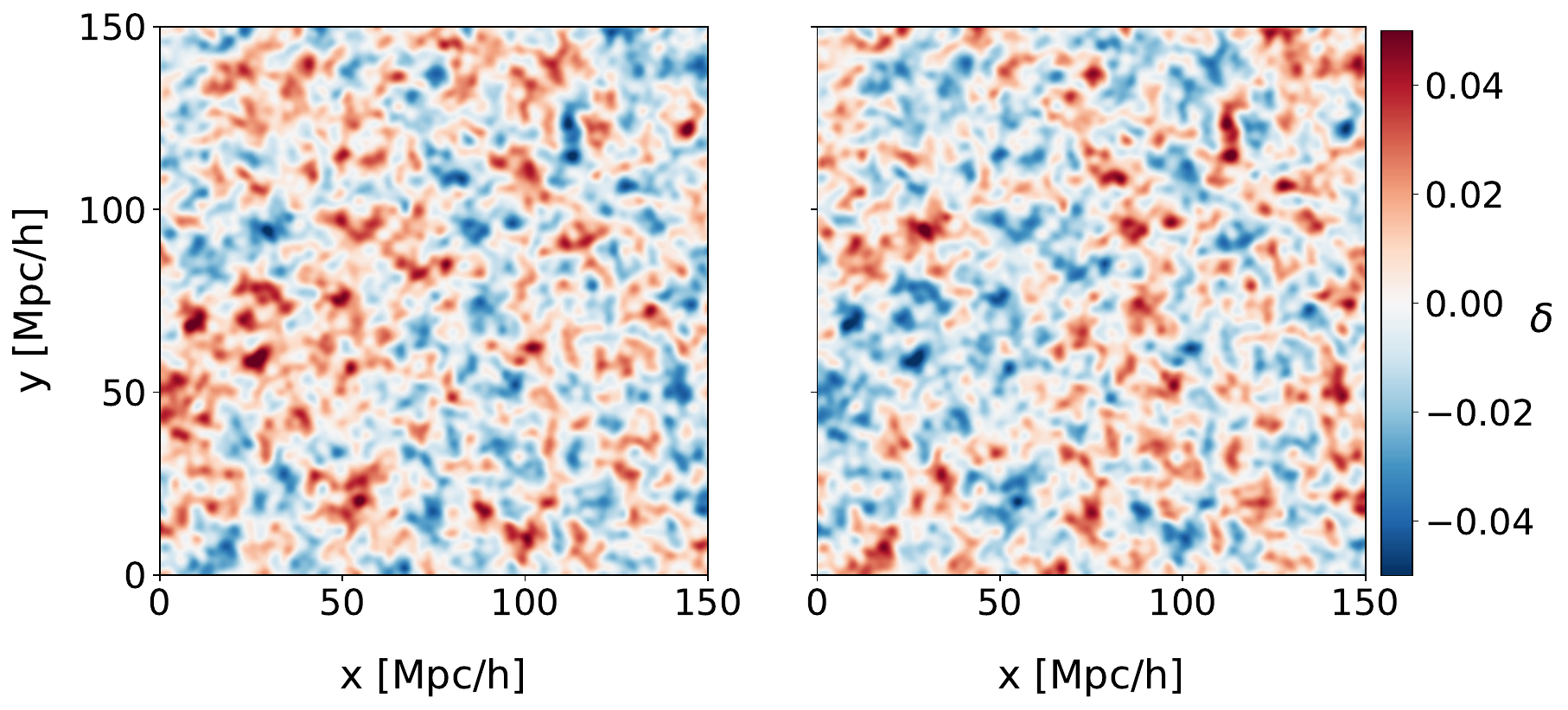}
    \caption{Initial density fields smoothed with a Gaussian over $2$\mpcph\ in the Mirror ({\it left}) and Reference ({\it right}) simulations at $z_{\rm i}=99$. Overdense (underdense) structures are colored in red (blue). Sign-inverted density fluctuations with identical shape and amplitude are shown.}
    \label{fig:InitDenMap}
\end{figure}

Among the Multiverse simulations, there is a pair of simulations that adopt the WMAP 5--year cosmology \citep{dunkley+09}. They are designed to relate the behavior of underdense and overdense regions compared with their counterparts (overdense and underdense regions, respectively). The GOTPM code employs the Fourier approach to generate the initial conditions. With a given set of random numbers, it assigns the amplitude and phase to each Fourier mode. Using these generated modes, the GOTPM code calculates the real-space density field through the Fast Fourier transformation. To generate the initial conditions for the Mirror simulation, we shift the phase of every Fourier mode of the Reference simulation by 180 degrees to flip the sign of all modes, namely $\delta_{{\rm R}}(x,t_{i})=-\delta_{{\rm M}}(x,t_{i})$. An overdense region in the Mirror simulation, for example, becomes an underdense region in the Reference simulation in the initial conditions.

The second-order Lagrangian Perturbation Theory \citep[2LPT;][]{jenkins+10} is applied to obtain the initial displacement and velocity of the simulated dark matter particles. Each particle carries the identification (ID) number which will be used to not only trace the particle evolution but also find its counterpart particle in the Mirror universe.

A subset of the two initial density fields constructed from the particle distributions at $z_{\rm i}=99$ is illustrated in Figure~\ref{fig:InitDenMap}. The density is calculated at the center of each pixel with a size ($1h^{-1}$Mpc)$^{3}$ using the cloud-in-cell (CIC) assignment scheme and smoothed with a Gaussian filter over $2$\mpcph. The excess (deficit) in density relative to the cosmic mean is colored in red (blue). Overdense structures in the Mirror simulation become underdense counterparts with equal amplitude at the same position in the Reference simulation and vice versa.

We evolve the particles in the WMAP 5-year $\Lambda$CDM universe to $z=0$. Using the standard Friend-of-Friend (FoF) algorithm, we identified virialized dark matter halos in both simulations with the linking length $l_{\rm link}=0.2\left< l_p\right>$, where $\left<l_p\right>$ is the mean particle separation of the simulations.

\subsection{Voids as cluster counterparts}
To identify voids in the Reference simulation, we use the cluster-scale dark matter halos in the Mirror simulation. We trace the particles at redshifts $z=0,1$, and $99$ in the Reference simulation that are the members of the clusters with  $M\ge10^{13}\Msun$ at $z=0$ in the Mirror simulation. There are $422,818$ such clusters. The cluster member particles in the Mirror simulation are called void particles in the Reference simulation. The cluster-counterpart voids are defined by these void particles. Thus, the cluster-counterpart voids in the Reference simulation are the underdense regions that would have evolved into massive clusters in the Mirror simulation. Note again that our void identification only relies on the membership of particles in the Mirror simulation.

\begin{figure}
	\includegraphics[width=\columnwidth]{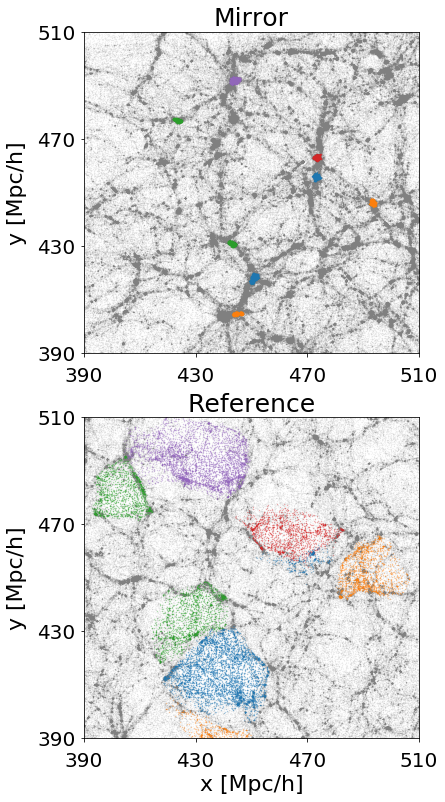}
    \caption{An example of particle distributions in the Mirror ({\it top}) and Reference ({\it bottom}) simulations at $z=0$. The member particles of eight clusters in the Mirror simulation and their corresponding void particles in the Reference simulation are marked with colors.}
    \label{fig:PTdistri}
\end{figure}

To illustrate the concept, the projected particle distributions of the Mirror (top) and Reference (bottom) simulations at $z=0$ are compared in Figure~\ref{fig:PTdistri}. We display every fifth particle in a $20$\mpcph-thick slab. The colored particles in the Mirror and Reference simulations represent the member particles of clusters with $M\ge 10^{14}$\msolph\ and their corresponding void particles, respectively. In general, the regions dominated by overdense structures in the Mirror simulation, e.g. clusters, appear as regions of sparse particle distribution in the Reference simulation.

The eight voids in the Reference simulation are located close to the positions of the corresponding same-color clusters in the Mirror simulation. In the Reference simulation, the void particles spread over a large volume. We often find clumps of particles at the boundaries of the voids. This is consistent with the formation of the ``boundary ridge'' of voids predicted in the spherical expansion model \citep{filmore&goldreich84,suto+84,bertschinger85,dunkley+09}. The cluster-counterpart voids also manifest tenuous inner structures like filaments and walls within them as was expected in theory \citep{sahni+94} and found in observation \citep{kreckel+11,alpaslan+14} and simulation \citep{dubinski+93,weygaert&kampen93,gottlober+03,reider+13}. Such inner structures are remnants of the boundary collisions between smaller voids within a large void \citep{dubinski+93,sahni+94}.

To determine the volume of individual voids, we define void pixels. A pixel in the simulation box is defined as a void pixel if its nearest particle is a void particle. Consequently, each void pixel carries the corresponding cluster-ID in the Mirror simulation, which is used to distinguish between individual void regions. Our method does not resort to a particular choice of void shape and density threshold. By definition, there is no volume overlap among neighboring voids.

\begin{figure}
	\includegraphics[width=\columnwidth]{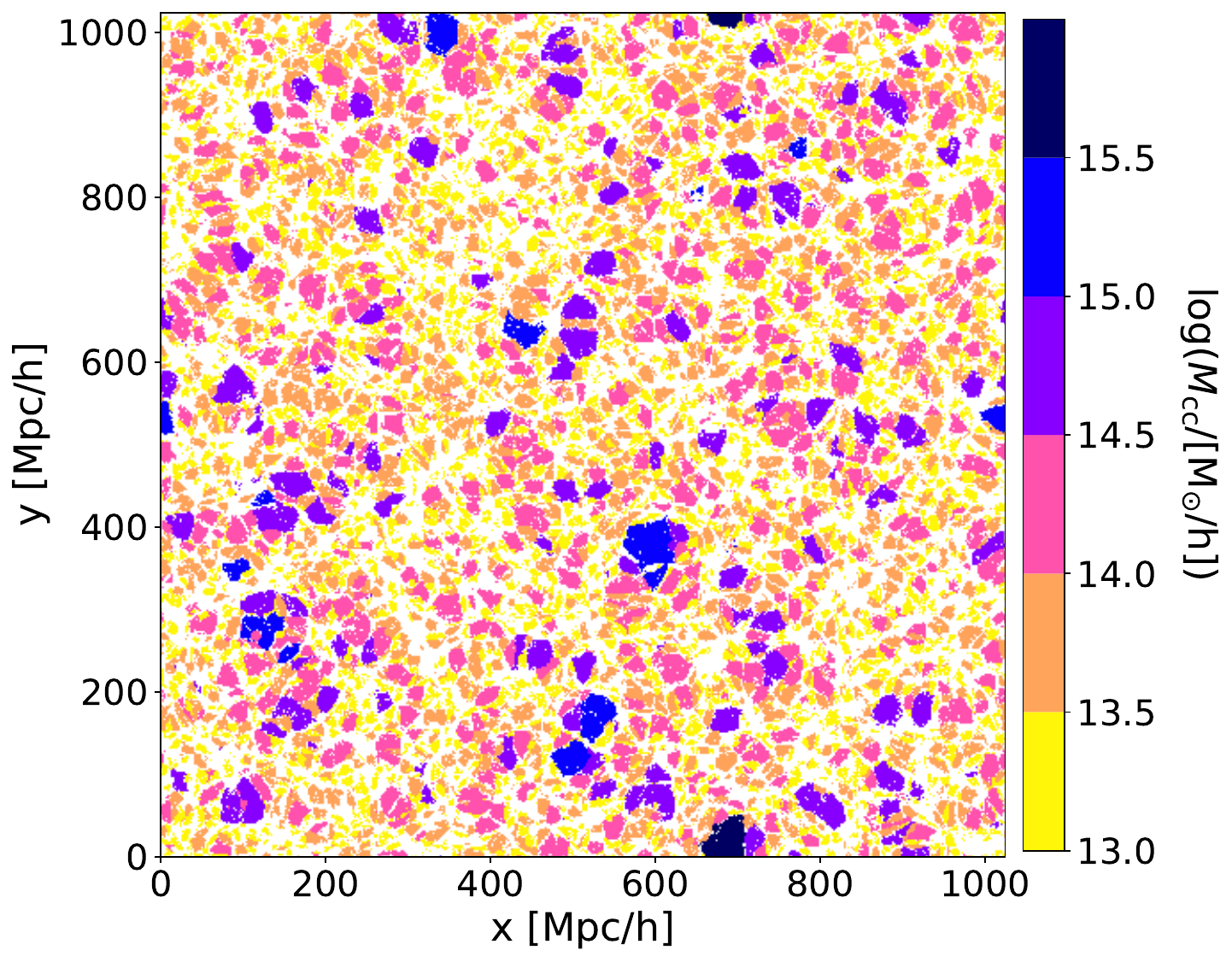}
    \caption{Spatial distribution of the voids corresponding to the mirror clusters in various mass bins above $10^{13}\Msun$. The thickness of the slab is $2$\mpcph. Simulated universe is mostly occupied by the voids.}
    \label{fig:voids}
\end{figure}

Figure~\ref{fig:voids} illustrates the spatial distribution of the voids corresponding to the clusters in the Mirror simulation, hereafter mirror clusters, color-coded according to the cluster mass bin. The figure clearly shows that most of the volume of the Reference universe is occupied by voids corresponding to the mirror clusters with $M_{\rm cc}\geq10^{13} \Msun$, where $M_{\rm cc}$ is corresponding mirror cluster mass. Roughly speaking, the physical sizes of voids increase with the mass of corresponding mirror clusters. From the visual inspection, the void shapes are aspherical in most cases. Hereafter, whenever we mention the ``corresponding mass'' of a void, it is the mass of its corresponding cluster in the Mirror simulation.

\section{Properties of voids}\label{sec:result}
\subsection{Density profile of voids}\label{subsec:profile}
Since one of the important void properties is its radial density profile, we examine the mass distribution within and around the voids. We first compute the densities at $1h^{-1}$Mpc-size regular-grid pixels using the spline kernel \citep{monaghan&lattanzio85} with a variable smoothing scale enclosing the same mass. We set the smoothing scale equivalent to the distance to the twentieth nearest particle from a pixel center. The density profile can be calculated either using all pixels or using only void pixels.

We calculate the density profile by averaging the densities at the pixels within concentric shells with inner and outer radii $[r,r+dr)$ from the geometric center of the voids. To stack the profiles of voids in different corresponding mass bins, the distance $r$ is normalized by the effective radius of each void,
\begin{equation}
    r_{{\rm v}}\equiv\left(\frac{3V_{\rm v}}{4\pi}\right)^{\frac{1}{3}},
	\label{eq:reff}
\end{equation}
where $V_{\rm v}$ is the void volume calculated by summing all the member void pixels.

\begin{figure}
	\includegraphics[width=\columnwidth]{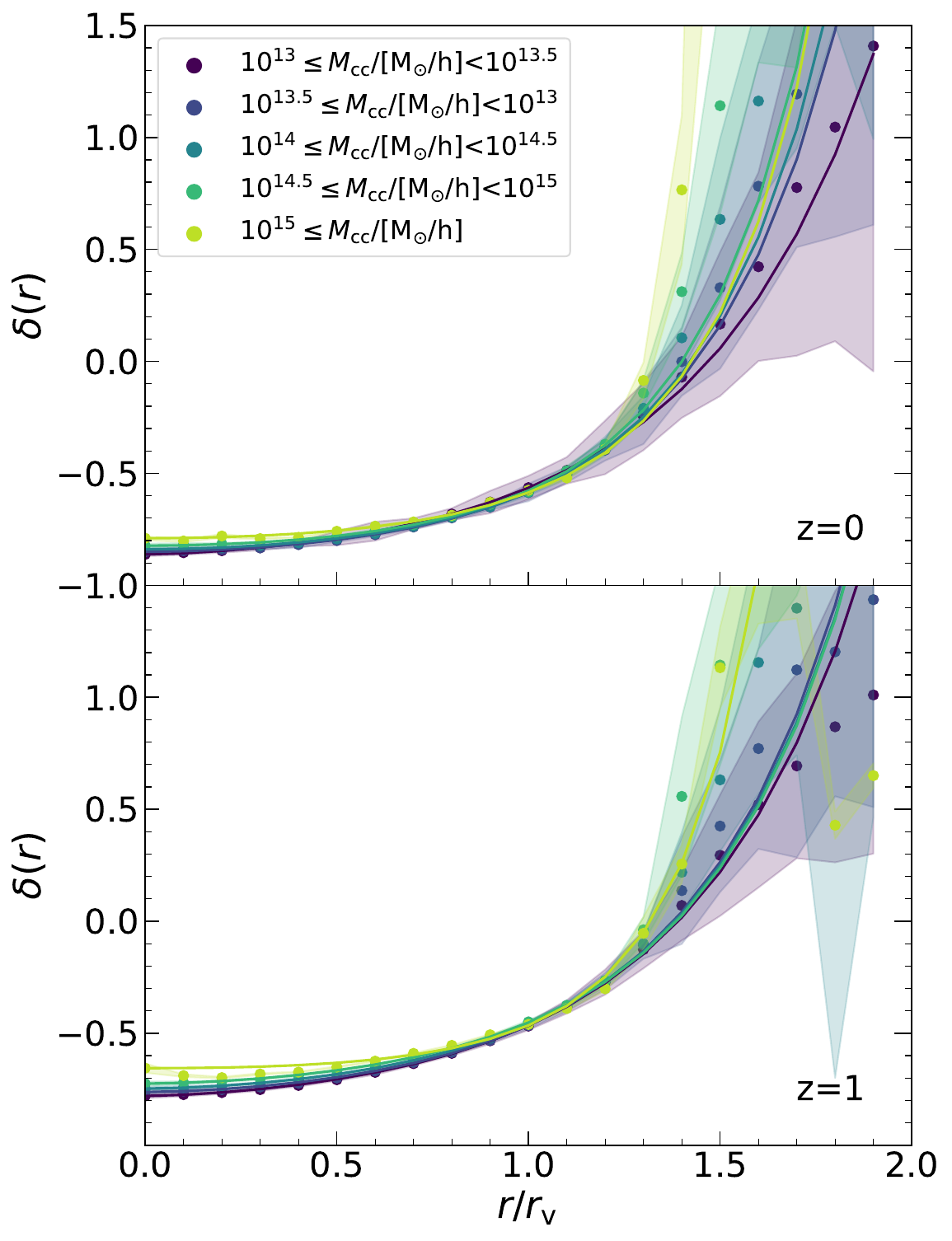}
    \caption{Void density profiles for different corresponding mass bins measured using the void pixels only. Shaded regions represent the standard deviations scaled down by $20$ for visibility. Solid lines are the best-fits with the universal form given in Eqn.~(\ref{eqn:vprof}). The distance to a shell $r$ is normalized by the effective radius of each void.}
    \label{fig:denprof_vpix}
\end{figure}

Figure~\ref{fig:denprof_vpix} shows the void density profiles for various corresponding mass bins measured with void pixels at different redshifts. We detect the flat underdensity floor at the central part within $0.5r_{\rm v}$ and steeply rising density wall at $r\gtrsim1.3 r_{{\rm v}}$. Voids have similar density distribution within $r_{\rm v}$ although the density at the innermost region tends to be slightly higher for a void with a larger corresponding mass. On the other hand, the profiles deviate from each other at a large radius range $r\gtrsim1.3 r_{\rm v}$. In this range, the slope becomes steeper for a void with a larger corresponding mass. Since a more spherical void tends to have a steeper slope at $r_{\rm v}$, our finding is consistent with the previous finding that larger voids tend to be more spherical \citep{park&lee07}. 
Given that voids are not perfectly spherical \citep{platen+08}, the profile deviations at outskirts ($r\ge r_{\rm v}$), in particular, can possibly disappear if one takes individual void shapes into account when computing void density profiles \citep{cautun+16}.

Also, we observe the redshift evolution of the void profiles. For a given corresponding mass bin, a density profile at $z=0$ is lower than that at $z=1$ below $r\simeq1.3 r_{{\rm v}}$. Hence, we find that the inner region becomes more underdense because the matter is continuously evacuating from the center. The void mass distribution and its evolution are consistent with the prediction in the spherical expansion model \citep{filmore&goldreich84,dubinski+93,sheth&weygaert04}. It is worthwhile to note that we find a universal function that well describes the density profiles up to $r\simeq 1.2r_{\rm v}$. Motivated by \cite{weygaert&kampen93} and \cite{colberg+05}, we propose an exponential function as
\begin{equation}
    \delta(r)=\delta_0+a_1 \left[{\rm exp}\left(\frac{r}{r_{\rm v}} \right)^{a_2} -1\right],
	\label{eqn:vprof}
\end{equation}
where $\delta_0$ is the minimum density at the void center ($r=0$). We provide the coefficients of the best-fitting lines for different corresponding mass bins in Table~\ref{tab:vpixprof_fitting}.

\begin{figure}
	\includegraphics[width=\columnwidth]{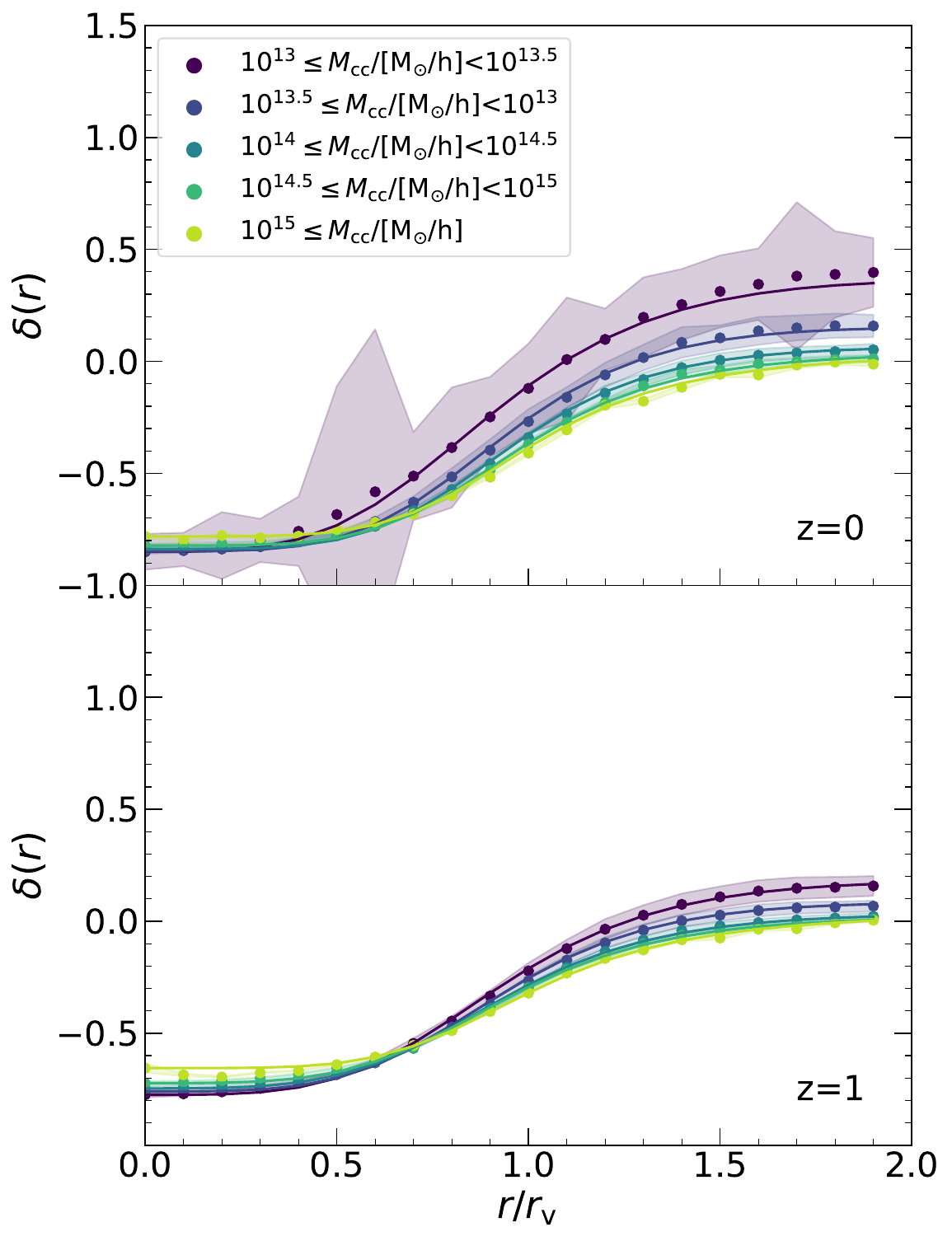}
	\caption{Same as Fig.~\ref{fig:denprof_vpix} but calculated using all pixels. In this case, solid lines represent the best fitting profiles obtained by adopting the universal form provided in \citet{hamaus+14}.}
	\label{fig:denprof_allpix}
\end{figure}

In Figure~\ref{fig:denprof_allpix}, we plot the density profiles measured by using all pixels. In the inner void region below $\sim0.5r_{\rm v}$, the density profile is consistent with the previous void-pixel case. We again find the minimum density of a void generally increases with its corresponding mass. Since a larger void tends to have larger corresponding mass, this trend qualitatively agrees with the results of \citet{hamaus+14} and \citet{sutter+14a} who reported a higher central density for a larger void.

A remarkable difference can be observed compared to Figure~\ref{fig:denprof_vpix} that the density at the outskirt converges to a finite density level. This is because of the dominating contributions from non-void pixels at $r \gtrsim r_{\rm v}$. We find a dependence of the outskirt densities on the corresponding mass. This is comparable with the argument in previous literature that smaller voids develop denser boundary ridge \citep{ceccarelli+13,hamaus+14,sutter+14a}. Note that the all-pixel void profiles are well described by the empirical formula suggested in \cite{hamaus+14}.

\begin{table}
	\centering
	\caption{Minimum densities and the best-fit coefficients of the universal function in Eqn.~(\ref{eqn:vprof}) for various corresponding mass bins at $z=0$ and 1.}
	\label{tab:vpixprof_fitting}
	\begin{tabular}{clccc} 
		\hline\hline
		redshift & log$_{10}(M_{\rm cc}/[\Msun])$ & $\delta_0$ & $a_1$ & $a_2$\\
		\hline
		$z=0$ & $[13,13.5)$ & $-0.862$ & $0.173$ & $1.51$\\
		& $[13.5,14)$ & $-0.850$ & $0.153$ & $1.74$\\
		& $[14,14.5)$ & $-0.839$ & $0.147$ & $1.81$\\
		& $[14.5,15)$ & $-0.823$ & $0.143$ & $1.92$\\
		& $[15,\infty)$ & $-0.790$ & $0.119$ & $2.00$\\ \hline 
		$z=1$ & $[13,13.5)$ & $-0.780$ & $0.183$ & $1.54$\\
		& $[13.5,14)$ & $-0.762$ & $0.173$ & $1.63$\\
		& $[14,14.5)$ & $-0.747$ & $0.168$ & $1.63$\\
		& $[14.5,15)$ & $-0.725$ & $0.159$ & $1.65$\\
		& $[15,\infty)$ & $-0.656$ & $0.111$ & $2.37$\\
		\hline
	\end{tabular}
\end{table}

\subsection{Physical size of voids}\label{subsec:size}
We now examine a one-to-one correspondence between the physical properties of the voids and corresponding mirror clusters. We examine the volume fraction and size of voids as a function of corresponding mass. By visually inspecting Figure~\ref{fig:voids}, one can expect that a massive mirror cluster is likely to be mapped to a large void in the Reference universe. Because all particles are placed on a regular grid in the ``pre-initial'' conditions, the initial volume of a void linearly increases with its number of void particles and thus with the corresponding mass of a mirror cluster.

\begin{figure}
	\includegraphics[width=\columnwidth]{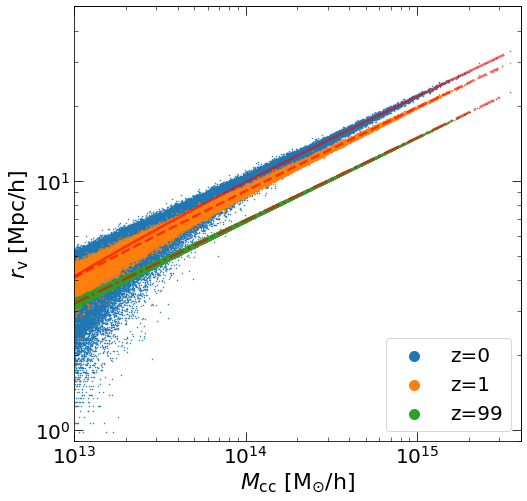}
    \caption{Scatter plot of $422,818$ void-cluster pairs showing a power-law scaling relation between the effective radius of a void, $r_{\rm v}$ and the corresponding cluster mass, $M_{\rm cc}$. The fitting functions of the scaling relation at $z=0$ (solid), $z=1$ (dashed), and $z=99$ (dot dashed) are presented in red.}
    \label{fig:sizemass}
\end{figure}

Figure~\ref{fig:sizemass} shows the relation between the comoving effective radii of voids and their corresponding mass at different redshifts. As expected, the effective radius of a void increases as a power law with the corresponding mass. In particular, the correspondence becomes tighter in the higher mass range at all redshifts. However, the scatter in the void radius increases in the small mass range ($M_{\rm cc}\le10^{14}$ \msolph). The increasing scatter of the relation at smaller masses implies that smaller voids are more vulnerable to the squeezing of the surrounding overdense region \citep{sheth&weygaert04, hamaus+14}. Considering that the scatter becomes large at $z=0$, we suggest that this non-linear evolution of squeezing only recently has become effective.

We also derive a fitting function of the scaling relation which is well represented by a power-law form as
\begin{equation}
    \left(\frac{r_{{\rm v}}}{h^{-1}{\rm Mpc}}\right)=c_1\cdot \left(\frac{M_{\rm cc}}{\Msun}\right)^{c_2}+ c_3,
	\label{eq:scaling}
\end{equation}
with dimensionless coefficients. The fitting coefficients for different redshifts are provided in Table~\ref{tab:coeff}. We note the redshift dependence of the coefficients. In principle, the initial ($z=99$) value of the exponent in the fitting function should be close to $1/3$. This is because the effective radius of a void initially follows the relation $r_{\rm v}\sim M_{\rm cc}^{1/3}$. We find that $c_2$ monotonically increases with redshift and matches the predicted value at $z=99$. On the other hand, $c_1$ increases with decreasing redshift because voids continually expand. Conversely, $c_2$ and $c_3$ decrease because of the recent squeezing of small corresponding mass voids, which results in smaller void radius.

\begin{table}
	\centering
	\caption{Fitting coefficients of the scaling relation between the void radius and corresponding mass.}
	\label{tab:coeff}
	\begin{tabular}{lccc} 
		\hline \hline
		redshift & $c_1$ & $c_2$ & $c_3$\\
		\hline
		$z=0$ & $4.09\times10^{-4}$ & $3.17\times10^{-1}$ & $-1.22$\\
		$z=1$ & $2.69\times10^{-4}$ & $3.25\times10^{-1}$ & $-4.11\times10^{-1}$\\
		$z=99$ & $1.48\times10^{-4}$ & $3.33\times10^{-1}$ & $1.14\times10^{-3}$\\
		\hline
	\end{tabular}
\end{table}

\begin{figure}
	\includegraphics[width=\columnwidth]{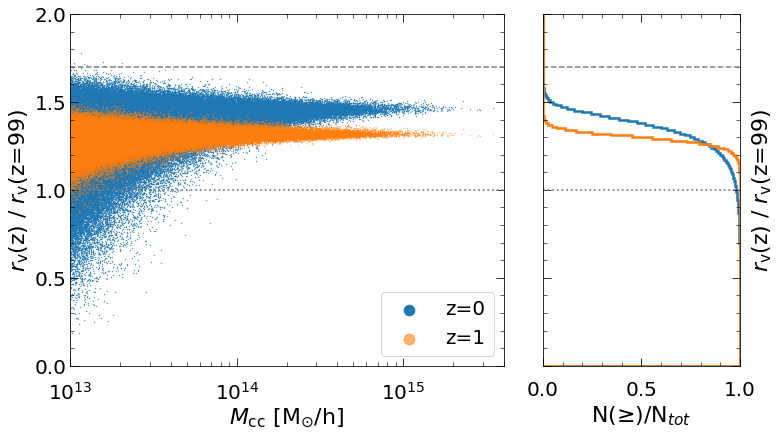}
    \caption{{\it Left}: Scatter plot of the relative size of voids at $z=0$ and 1 compared to $z=99$. Voids with ratio below unity (grey dotted) and above $1.7$ (grey dashed) have experienced squeezing and shell-crossing, respectively. {\it Right}: Cumulative number fractions (horizontal) of voids with a ratio above a particular value (vertical).}
    \label{fig:sizeevol}
\end{figure}

We now focus on the size evolution of voids and show their relative growth in the left panel of Figure~\ref{fig:sizeevol}. To quantify the relative growth of voids, we compute the ratio of a void radius at a particular redshift over that at $z=99$. A ratio above unity indicates that voids have expanded. In general, most of the voids expand as they evolve. Note that few cluster-counterpart voids reach the shell-crossing at which the predicted radius ratio is $\sim1.7$ for spherical expansion model voids. The right panel shows the cumulative fraction of the voids with radius ratio greater than a certain value. The fraction of voids smaller than their initial size is about $0.03\%$ at $z=1$ but increases to $2\%$ at $z=0$. This supports our previous claim that the compression on voids by larger surrounding overdensities has become effective only recently. It also agrees with the previous result of \citet{pontzen+16} who showed the average density evolution of anti-halos. They found some of the counterparts of small mass halos become overdense structures due to the squeezing at late time. Note that the fraction of contracted voids may increase when we include smaller voids corresponding to the clusters with $M_{\rm cc}\le10^{13}$\msolph.

\begin{figure}
	\includegraphics[width=\columnwidth]{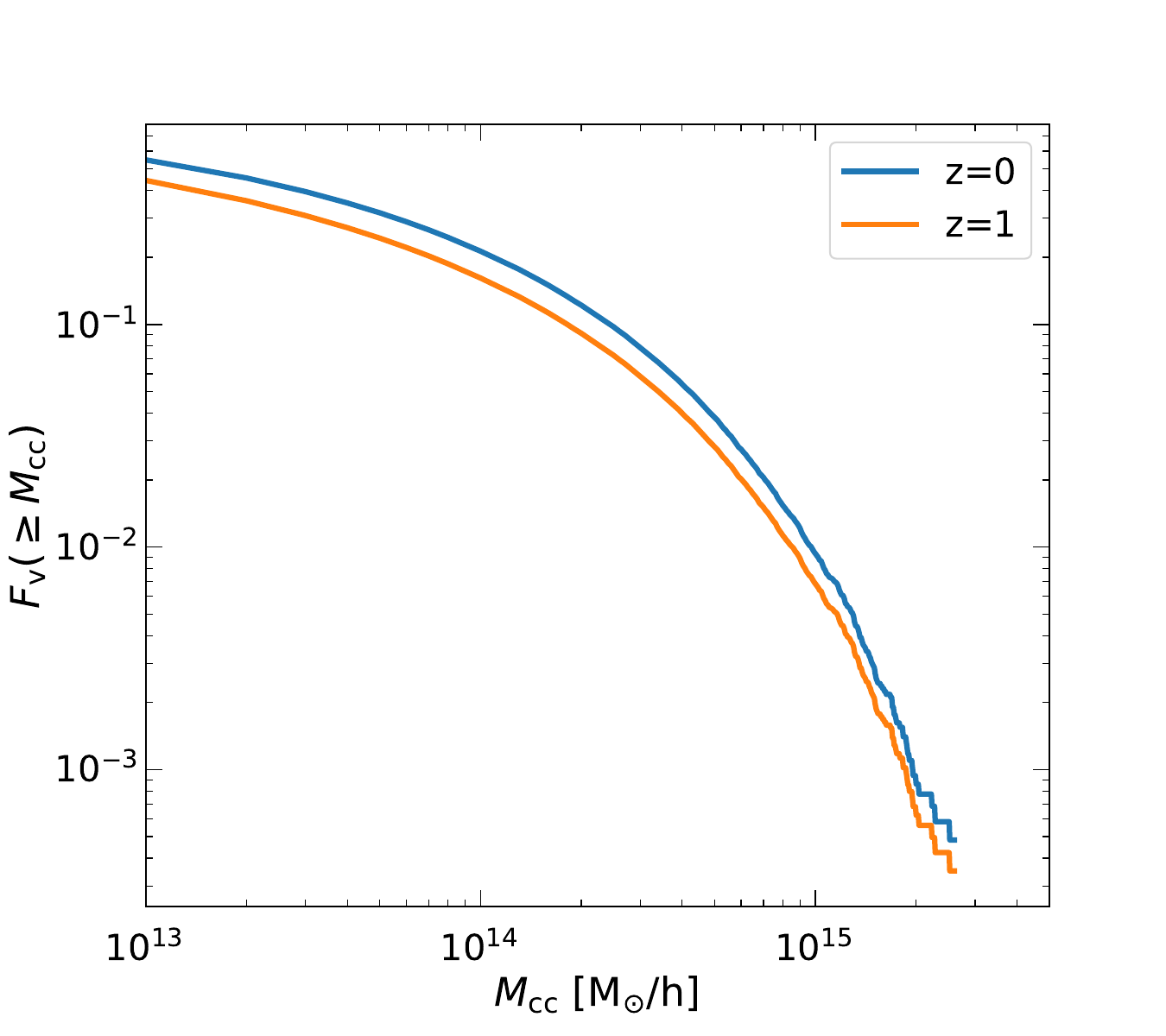}
    \caption{Cumulative void volume fraction as a function of corresponding cluster mass.}
    \label{fig:massfrac}
\end{figure}

\subsection{Volume fraction of voids}\label{subsec:volfrac}
Now we examine the void volume fraction $F_{\rm v}$ defined as
\begin{equation}
    F_{\rm v} = \frac{\sum_i V_{\rm v}^i}{V_{\rm box}},
\end{equation}
where $V_{\rm v}^i$ is the volume of $i$'th void and $V_{\rm box}$ is the entire simulation volume. Figure~\ref{fig:massfrac} shows the cumulative volume fraction of voids with corresponding mass more massive than $M_{\rm cc}$. The exponential drop at a high mass range shows that most of the volume is occupied by the voids with small corresponding mass.
This is qualitatively consistent with the result of \citet{furlanetto&piran06} that large voids in the galaxy distribution fill only a small fraction of the volume of the universe.

On the other hand, the volume fraction reaches $F_{\rm v}\simeq0.54$ for voids with corresponding mass above $10^{13}$\msolph. These voids fill more than $40\%$ of the universe at $z=1$. The comparison between the two redshifts clearly shows the redshift evolution of the volume fraction or the status of void expansion on average. Although the volume fraction of our voids changes with corresponding mass, it generally agrees with the values ($0.08\le F_{\rm v} \le1$) obtained for various void finders \citep{colberg+08,cautun+14}. The large variation of $F_{\rm v}$ for voids identified with those finders may imply that their corresponding mass ranges differ depending on the void-finding algorithm.

The volume fraction of the voids as a function of corresponding mass provides us a clue to the lowest mass-scale of voids we can detect in practice, for example, from a density field. From the fact that voids are likely to percolate at high volume fractions \citep{shandarin+06}, identifying individual voids with lower corresponding mass-scale around $10^{13}\Msun$ will be a challenging task.

\section{Identifying voids from density fields}\label{sec:discussion}
Because we are not able to observe our mirror Universe, one may wonder whether it is possible to find voids without proper information on their counterparts in the mirror universe. Hence, we develop a method to identify cluster-counterpart voids directly from a given density field exploiting the relations found in the previous sections.

\begin{figure}
	\includegraphics[width=\columnwidth]{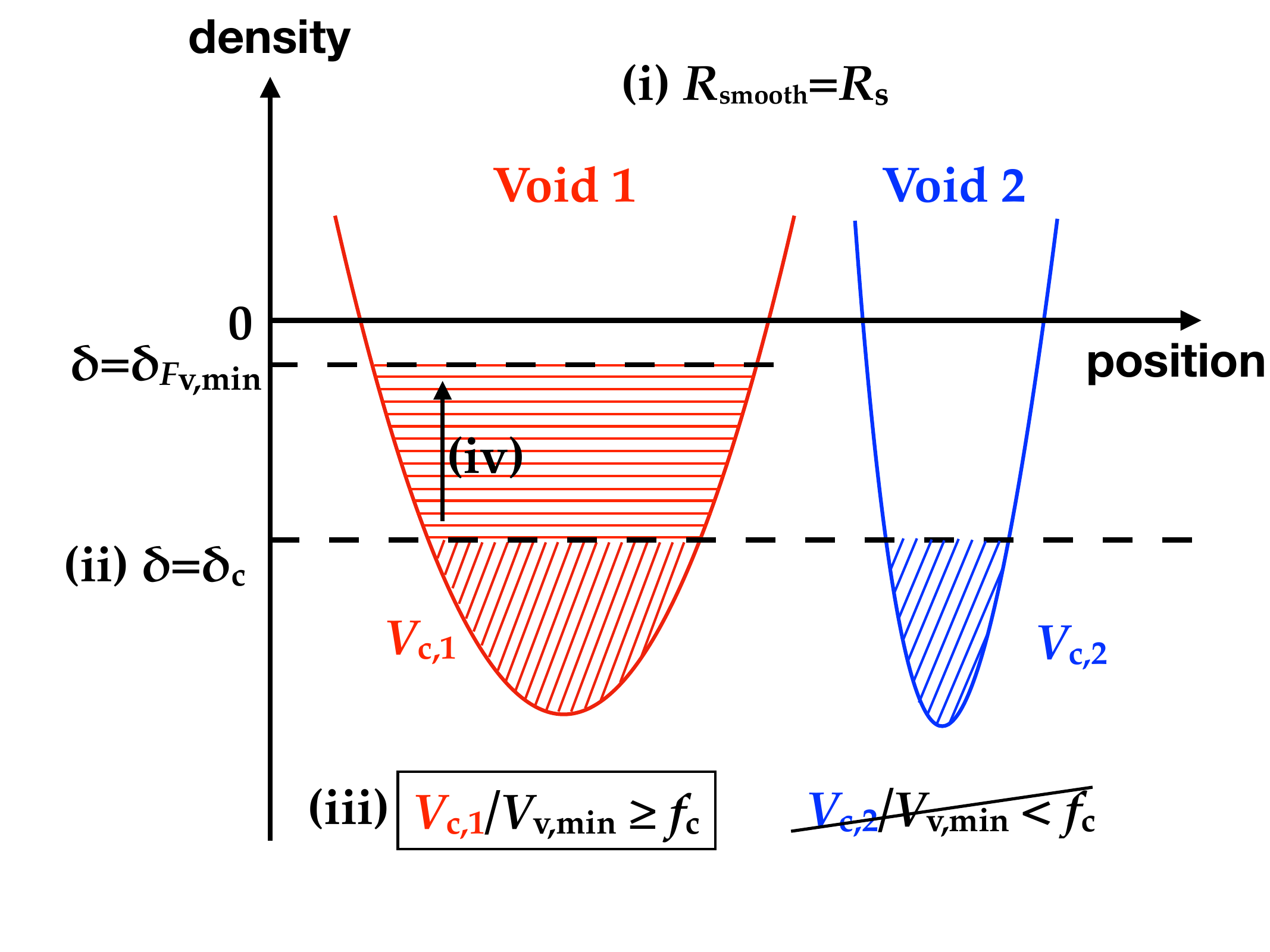}
    \caption{An illustration of our method to recover voids corresponding to clusters more massive than $M_{\rm min}$. Density distributions of Void 1 ($M_{\rm cc,1}\ge M_{\rm min}$) and Void 2 ($M_{\rm cc,2}<M_{\rm min}$) are schematically depicted. The sequence of the procedures: (i) smooth a density field with smoothing scale $R_{\rm s}$, (ii) identify underdensities (diagonally-hatched) below $\delta=\delta_{\rm c}$, (iii) discard cores (blue hatched) smaller than $f_{\rm c}V_{\rm v,min}$, and (iv) repeatedly connect next lowest-density volume elements to the remaining cores (red diagonally-hatched) to recover complete void volume (red hatched).}
    \label{fig:summary}
\end{figure}

\begin{figure*}
	\includegraphics[width=2\columnwidth]{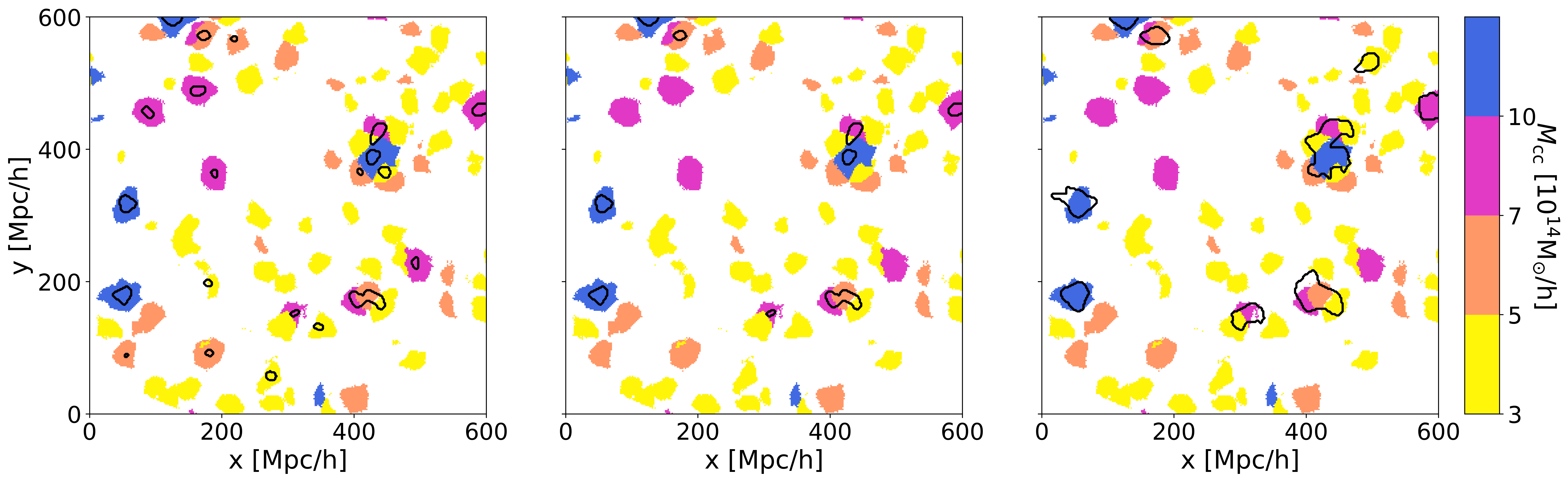}
    \caption{Spatial distribution of cores (black contours), the reference voids (blue), and other cluster-counterpart voids (non-black) in a $32$\mpcph-thick slice. We adopt the optimal values of the three parameters to identify reference voids with $M_{\rm cc}\ge 10^{15}$\msolph. We build the density field from the Reference simulation particles using the CIC method on the regular grids with 2 $h^{-1}$ Mpc spacing and smooth this CIC density field with the Gaussian kernel. {\it Left}: All underdensities below a given density threshold are identified as cores. {\it Middle}: Remaining cores after discarding smaller ones below $V_{\rm c}= f_{\rm c}V_{\rm v,min}$. {\it Right}: Voids finally identified by expanding the remaining cores. For more details, see text.}
    \label{fig:udNvoid}
\end{figure*}

\subsection{Strategy of void identification}\label{sub:strategy}
According to our discovery of the void size - cluster mass relation, a larger void monotonically corresponds to a more massive cluster. In addition, the universal density profile of voids implies that an underdense region below a given density threshold, hereafter a void core, is larger for a larger void. Hence, to identify voids corresponding to clusters more massive than $M_{\rm min}$, we begin by finding void cores larger than a typical core volume of the void with $M_{\rm cc}=M_{\rm min}$. We then expand those void cores to recover the expected total volume of the voids.

Figure~\ref{fig:summary} schematically shows our method of identifying voids with $M_{\rm cc}\ge M_{\rm min}$ by utilizing three parameters: smoothing scale $R_{\rm s}$, density threshold $\delta_{\rm c}$, and minimum core fraction $f_{\rm c}$. We (i) start with the density field smoothed over $R_{\rm smooth}=R_{\rm s}$ with the Gaussian filter, and (ii) identify void cores (diagonally-hatched) as the underdense regions below $\delta=\delta_{\rm c}$. We then (iii) select the cores with $V_{\rm c}\ge f_{c}V_{\rm v,min}$ as the cores (red diagonally-hatched) of the voids in a desired corresponding mass range, where $V_{\rm c}$ and $V_{\rm v,min}$ are the volume of void cores and the expected volume of the void with $M_{\rm cc}=M_{\rm min}$, respectively. To recover the full void volume (red hatched), we (iv) expand the selected cores by attaching adjacent lowest-density regions to the remaining cores similarly to the watershed method until their global volume fraction reaches $F_{\rm v,min}$ (see Figure \ref{fig:massfrac}), where $F_{\rm v,min}$ is the predicted volume fraction of the voids with $M_{\rm cc} \ge M_{\rm min}$.

Figure~\ref{fig:udNvoid} exemplifies how our method finds the voids corresponding to the clusters more massive than $M_{\rm min}=10^{15}\Msun$. Hereafter, we call the cluster-counterpart voids defined by the void pixels corresponding to a mass range $M_{\rm cc}\ge M_{\rm min}$ ``reference voids''. In Figure~\ref{fig:udNvoid}, we show the reference voids and the regions identified by using our method. The left panel shows the distribution of the cluster-counterpart voids (non-black) identified according to the void-cluster correspondence model while the black contours are the void cores found with $R_{\rm s}=7.3$\mpcph, $\delta_{\rm c}=-0.716$ (the step (ii) in Figure~\ref{fig:summary}).

In the middle panel, we attempt to extract the cores of the reference voids by deleting the core regions smaller than $f_{\rm c}V_{\rm v,min}$ (the step (iii) in Figure~\ref{fig:summary}). We adopt $f_{\rm c}=0.05$ for this plot. Note that some of the cores within smaller voids remain unremoved. This happens when a group of small voids forms a large void complex behaving similar to a single large void in a density field. Finally, to identify the complete void volume, we apply the watershed method around each remaining void core by raising the density threshold (the step (iv) in Figure~\ref{fig:summary}). The expanded black contours in the right panel of the figure mark the void regions identified in this way. In the following subsections, we will quantitatively measure the correspondence between the identified voids and the reference voids, and describe how we determine the optimal set of the three parameter values.

\subsection{Parameter dependence of void identification}\label{sub:compreli}
To investigate how the rate of correct void detection in our approach depends on the three parameters, we quantitatively measure the completeness and reliability 
of our void detection algorithm when the parameters are varied. We compute completeness and reliability defined as 

\begin{equation}
\mathcal{C} \equiv N_{\rm s}/N_{\rm v}
\end{equation}
and 
\begin{equation} \label{eq:reliability}
    \mathcal{R} \equiv N_{\rm s}/N_{\rm c},
\end{equation} 
respectively. Here, $N_{\rm v}$, $N_{\rm c}$, and $N_{\rm s}$ represent the number of reference voids, void cores, and successfully detected reference voids, respectively. The completeness is the fraction of the successfully detected reference voids whereas the reliability represents the fraction of the cores actually corresponding to the cluster-counterpart voids with $M_{\rm cc} \ge 0.8M_{\rm min}$. Here we adopt a lower mass limit of 80 \% of $M_{\rm min}$ because we want to allow the error in detecting void cores on a certain mass limit.

Now we explain how to couple a void core to a reference void. We calculate the distances between geometric centers of identified cores and cluster-counterpart voids and pair each core to its nearest void. A core--void pair is counted as a successful recovery of a reference void if the corresponding mass of the paired void satisfies the mass criterion. If the corresponding mass is lower than $M_{\rm min}$ (or $0.8M_{\rm min}$) but the core has several nearby voids, we pair the core to a void with the largest corresponding mass within $1.5r_{\rm c}$, where $r_{\rm c}$ denotes the effective radius of a core;
\begin{equation}
    r_c \equiv \left(\frac{3V_c}{4\pi}\right)^{1/3}.
\end{equation}
This is because the nearest void may not be the genuine corresponding void to a core when the core overlaps with multiple voids forming a void complex as shown in the middle panel of Figure~\ref{fig:udNvoid}. In this case, we assume that the void with the largest $M_{\rm cc}$ is the corresponding void of the core.

\begin{figure}
	\includegraphics[width=\columnwidth]{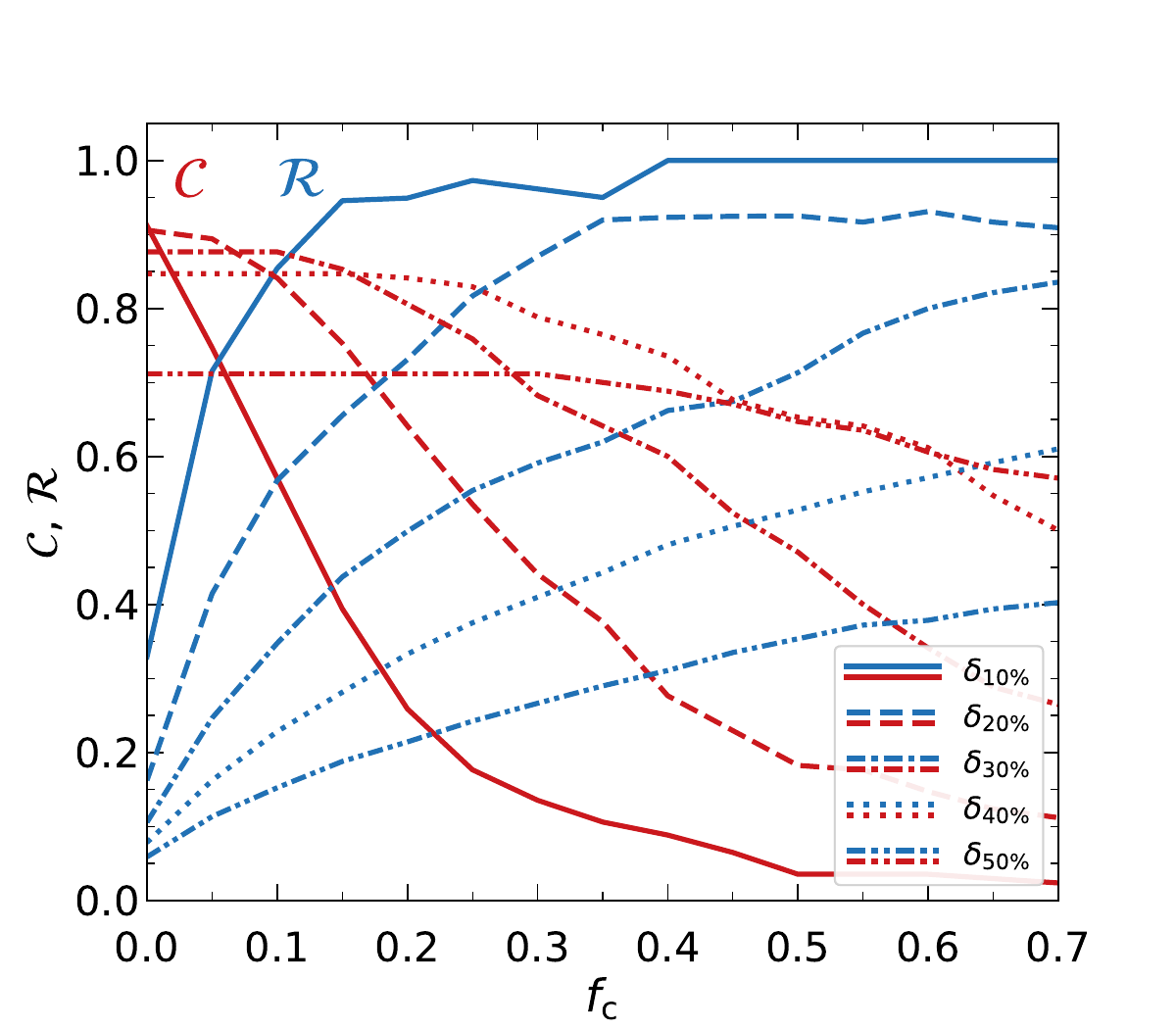}
    \caption{Completeness (red) and reliability (blue) of our void finding for the reference voids ($M_{\rm min}=10^{15}\Msun$) as a function of the minimum core fraction applying different density thresholds. We adopt the optimal smoothing scale $R_{\rm s}=7.3$\mpcph\ for the reference voids.}
    \label{fig:completeness_delta_vcutoff}
\end{figure}

Figure~\ref{fig:completeness_delta_vcutoff} shows the completeness (red) and reliability (blue) as a function of the minimum core fraction $f_{\rm c}$ when $M_{\rm min}=10^{15}$\msolph. We set $R_{s}=7.3$\mpcph\ here. We apply a density cut that defines the density-percentile volume within reference voids. We pixelate all reference voids and sort all the pixels in order of density, from which we know the percentile of the void volume below a density threshold. For example, $\delta_{10\%}$ means the density threshold detecting the most underdense 10\% volume of the reference voids.

We find that the completeness is a decreasing function of $f_{\rm c}$. This is because more cores are removed as we impose a larger $f_{\rm c}$. When $f_{\rm c}=0$, on the other hand, the completeness decreases with the increasing density threshold. This is because core regions percolate when a higher $\delta_{x\%}$ is applied. Also, the resulting center of a percolating core moves substantially. Consequently, the completeness drops. However, the reliability increases with $f_{\rm c}$. This implies that larger cores are more likely to correspond to the reference voids. On the other hand, for a given $f_{\rm c}$, the reliability declines with increasing $\delta_{x\%}$ because the number of cores increases more rapidly than that of successful void recoveries.

\begin{figure}
	\includegraphics[width=\columnwidth]{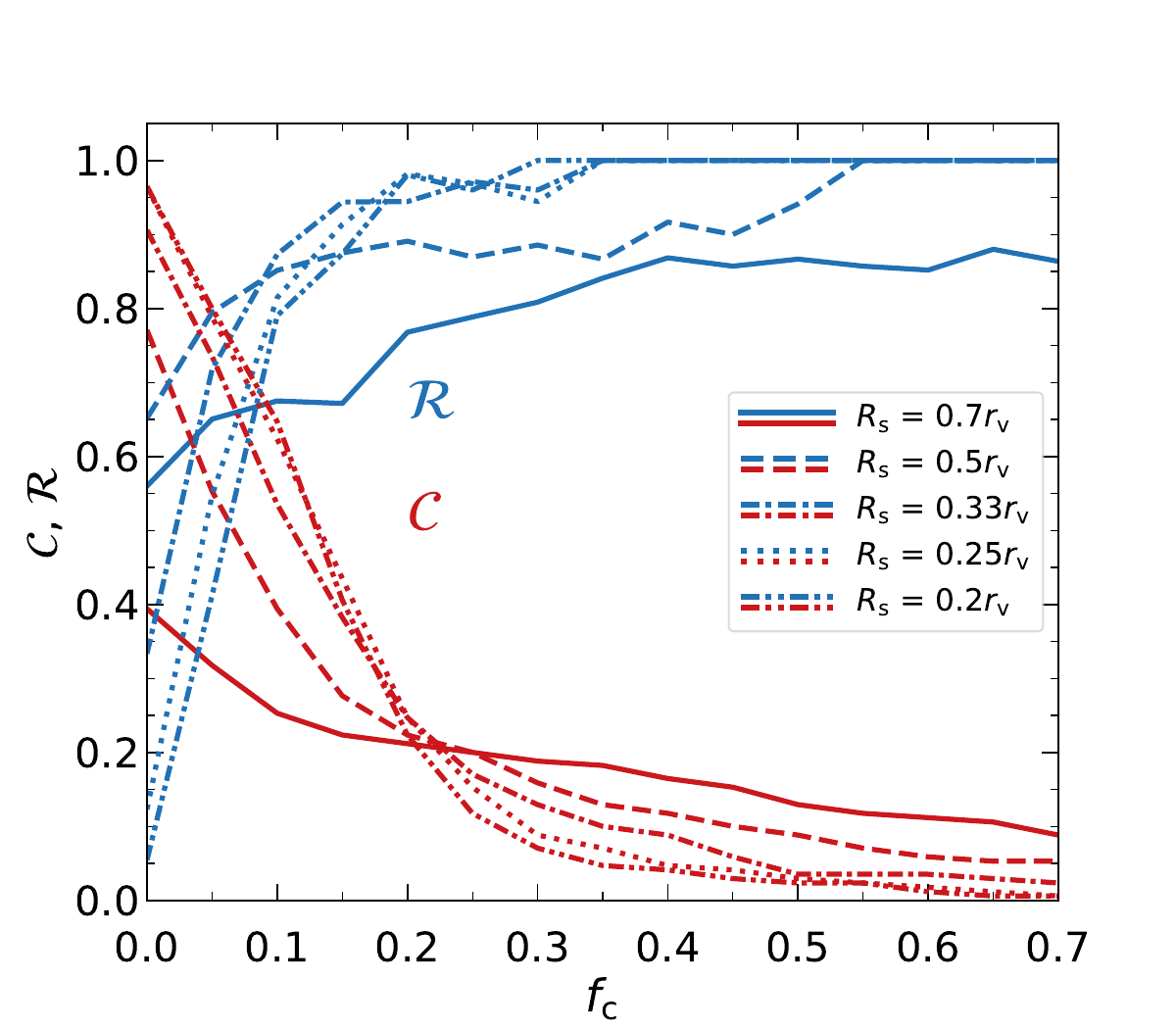}
    \caption{Completeness (red) and reliability (blue) of our void finding for the reference voids ($M_{\rm min}=10^{15}\Msun$) as a function of the minimum core fraction applying different smoothing-scales. We adopt the optimal density threshold $\delta=\delta_{10\%}$ for the reference voids.}
    \label{fig:completeness_scale}
\end{figure}

Now, we investigate the smoothing-scale dependence of completeness and reliability. We show the completeness (red) and reliability (blue) as functions of the minimum core fraction for various smoothing scales in Figure~\ref{fig:completeness_scale}. In this plot, the density threshold is fixed to $\delta_{\rm c}=\delta_{10\%}$. With smaller-scale smoothing, the completeness becomes higher for $f_{\rm c}<0.2$, whereas we obtain lower completeness for $f_{\rm c}\ge0.2$. The trend is the opposite in the reliability. This is a consequence that a single void core tends to be identified as several smaller cores when a density field is smoothed on a smaller scale. Hence, with a smaller smoothing length, the increased numbers of identified cores and successful detection yield higher completeness and lower reliability for $f_{\rm c}<0.2$. However, the completeness (reliability) becomes lower (higher) for $f_{\rm c}\ge0.2$ because the number of identified cores decreases more rapidly compared to the case of a larger-scale smoothing.

\subsection{Voids identified from a density field using the optimal parameter set}
\begin{table}
	\centering
	\caption{Optimal values of the parameters for identifying voids corresponding to various minimum cluster mass. In columns, the minimum corresponding cluster mass $M_{\rm min}$, completeness $\mathcal{C}$, reliability $\mathcal{R}$, smoothing scale $R_{\rm s}$, density threshold $\delta_{\rm c}$, and minimum core fraction $f_{\rm c}$ are provided.}
	\label{tab:params}
	\begin{tabular}{cccccc} 
		\hline\hline
		$M_{\rm min}$ [$10^{14}{\rm M}_{\odot}$/h]& $\mathcal{C}$ & $\mathcal{R}$ & $R_{\rm s}$ [Mpc/h]& $\delta_{\rm c}$ & $f_{\rm c}$\\
		\hline
		$3$ & 0.70 & 0.70 &  $4.8$ & $-0.746$ & $0.02$\\
		$5$ & 0.72 & 0.72 &  $5.7$ & $-0.734$ & $0.03$\\
		$7$ & 0.72 & 0.71 &  $6.4$ & $-0.727$ & $0.04$\\
		$10$ & 0.74 & 0.74 &  $7.3$ & $-0.716$ & $0.05$\\
		\hline
	\end{tabular}
\end{table}

To identify voids from a given density field in the most optimal way, we search for the set of parameter values which yields completeness and reliability as high as possible. Specifically, we select the parameter values of the highest balancing point between the completeness and reliability in the parameter space of $(R_{\rm s}, \delta_{x\%}, f_{\rm c})$. For example, for identifying reference voids ($M_{\rm min}=10^{15}\Msun$), we find the highest balance when $\delta_{\rm c}=\delta_{10\%}$ and $f_{\rm c}=0.05$ for $R_{\rm s}=7.3$\mpcph\ (Figure~\ref{fig:completeness_delta_vcutoff}). Similarly, for a given density threshold, we can find that the highest crossing happens when a density field is smoothed on the scale equal to $1/3$ of the effective radius of a void with $M_{\rm cc}=M_{\rm min}$ (Figure~\ref{fig:completeness_scale}). Thus, the optimal parameter values are $R_{s}=0.33r_{\rm v}$, $\delta_{\rm c}=\delta_{10\%}$, and $f_{\rm c}\simeq0.05$ for voids with $M_{\rm cc}\ge 10^{15}\Msun$.

We repeat similar analyses for different $M_{\rm min}$ and find that the optimal parameter values are robust against $M_{\rm min}$. The optimal density threshold and smoothing scale are $\delta_{10\%}$ and $0.33r_{\rm v}$, respectively. This self-similarity holds until $M_{\rm min}$ is lowered to $3\times10^{14}\Msun$. The optimal value of $f_{\rm c}$ slightly increases with $M_{\rm min}$. We summarize the optimal parameter values and resulting completeness and reliability for a different $M_{\rm min}$ in Table~\ref{tab:params}. Note that the optimal parameter values are determined solely by the corresponding mass-scale of reference voids. 

\begin{figure*}
	\includegraphics[width=2\columnwidth]{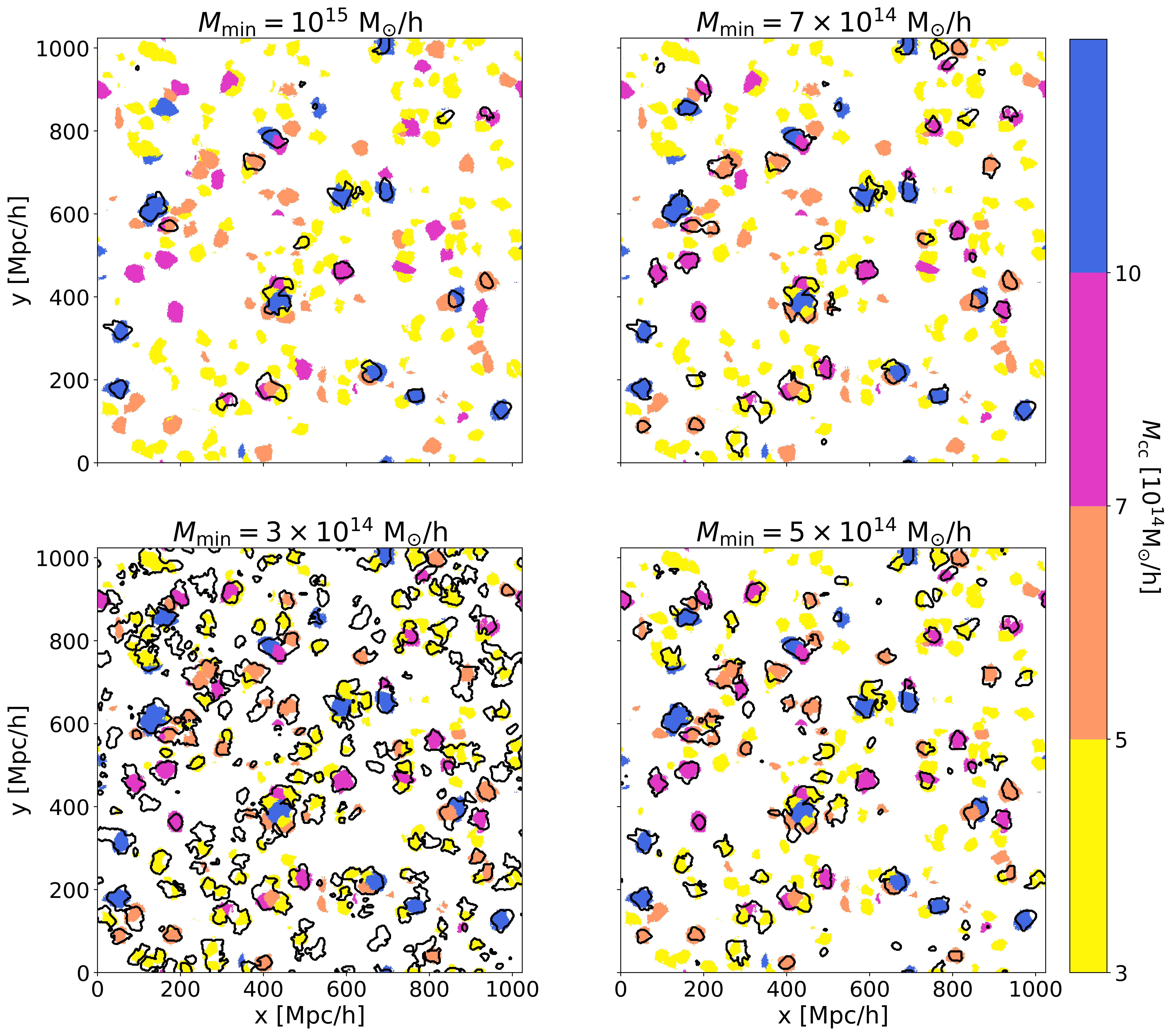}
    \caption{Spatial distribution of recovered voids (black contours) from density fields and cluster-counterpart voids (non-black) in a $32$\mpcph\ slice. Recovered voids with $M_{\rm min}=10^{15}\Msun, 7\times10^{15}\Msun, 5\times10^{15}\Msun$, and $3\times10^{15}\Msun$ are displayed in clockwise direction from top-left to bottom-left.}
    \label{fig:grownseeds}
\end{figure*}

Figure~\ref{fig:grownseeds} shows recovered voids (black contours) from density fields using this approach and cluster-counterpart voids (non-black) using its ``mirror'' information. Overall, recovered voids well coincide with their reference voids. For a smaller $M_{\rm min}$, the recovered voids are more likely to be identified as void complexes of neighboring cluster-counterpart voids. This phenomenon may be related to ``void hierarchy'' for which a large void encompasses several smaller subvoids \citep{weygaert&kampen93,sheth&weygaert04,neyrinck08,aragon-calvo+10,lavaux&wandelt12,aragon-calvo&szalay13}. Below $M_{\rm min}=3\times10^{14}\Msun$, it is almost impossible to distinguish individual voids because they mostly form large void complexes.

\section{Conclusions}\label{sec:conclusion}
We study the correspondence between cosmic voids and massive clusters using a pair of simulations with relatively inverted initial overdensity fields. We examine the physical properties of cluster-counterpart voids such as the radial density profile, size, and volume fraction, and develop a method to identify the voids corresponding to the clusters with mass above a particular limit, directly from a given density field. We identify voids in one simulation using void particles that are cluster-member particles in the mirror simulation.

We calculate void density profiles using $1)$ only void pixels and $2)$ all pixels. We find that both types of density profiles have their own universal forms. We observe that the outskirt ($r\ge1.3 r_{\rm v}$) density rises steeply when void pixels are used. In the all-pixel case, the density profile no longer diverges to high density at $r\ge1.3 r_{\rm v}$. We also confirm that the density at the void boundary decreases with void size \citep{hamaus+14,sutter+14a}. Overall, the density profiles of voids in the void-cluster model qualitatively agree with those of voids identified by void finders and predicted from the spherical expansion model.

We find a tight power-law scaling relation between the void size and the corresponding cluster mass. The scatter in the scaling relation is larger at low redshifts and for small voids. This indicates that smaller voids suffer more from squeezing and tunneling by surrounding environment than larger voids \citep{sheth&weygaert04}. However, we find that only about $2\%$ of voids at $z=0$ are smaller than their initial size ($z=99$) and that the squeezing becomes severe only recently ($z<1$). Volume fraction of voids increases from $1\%$ to $54\%$ as the corresponding mass cut changes from $M_{\rm cc} \gtrsim 10^{15}\Msun$ to $10^{13}\Msun$.

We develop a method to identify voids from a density field based on the findings that they have a universal density profile over a wide range of void size. We empirically determine the optimal parameter values that yield the highest completeness and reliability of our void finding. The optimal smoothing scale and density threshold are $0.33r_{\rm v}$ and $\delta_{10\%}$, and they are robust against the mass cut when $M_{\rm min} \gtrsim 3\times10^{14}\Msun$. The optimal value for $f_{\rm c}$ is within 0.02 and 0.05 and increases with the minimum corresponding mass. With these parameters, we achieve 70 -- 74 \% of completeness and reliability in recovering the reference-void cores corresponding to the clusters with mass limits from $3\times10^{14}$\msolph\ to $M_{\rm min}=10^{15}$\msolph. We reliably recover the complete void regions by raising the density threshold as in the watershed method until the void volume fraction reaches the reference $F_{\rm v,min}$ (see Figure~\ref{fig:massfrac}).

There are several interesting avenues to be pursued for further study. The immediate task is to test our algorithm in redshift space using biased tracers of mass density field, namely galaxies. Another avenue is to compare our method with existing void finders. Using the relation between void size and the corresponding mass-scale (see Figure~\ref{fig:sizemass}), we may detect the preferred void scale of each void finder and further fill the abundance gap between the theories and void finders \citep{chan+14,achitouv+15,nadathur&hotchkiss15}. On the other hand, we may be able to determine the parameter values of those void finders for the identification of voids that are equivalent to clusters of a particular mass. It is also interesting to examine the cosmology dependence of the void-cluster relation. In particular, in different gravity theories, not only the scaling relation between void radius and cluster mass will change but also their redshift evolution will deviate from $\Lambda$CDM. Thus, the redshift evolution of the scaling relation can be used for testing non-standard gravity theories.

\section*{Acknowledgements}
We thank an anonymous referee for helpful comments which helped improve the original manuscript. The authors thank S. Appleby for his helpful comments on this work. S.J.S. and C.B.P. were supported by KIAS Individual Grant PG071201 and PG016903 at Korea Institute for Advanced Study, respectively. J.H.K. was supported by a KIAS Individual Grant (KG039603) via the Center for Advanced Computation at Korea Institute for Advanced Study. The computing resources were kindly provided by the Center for Advanced Computation at Korea Institute for Advanced Study.


\end{document}